\documentclass[twocolumn,prl,amssymb]{revtex4}

\usepackage{times}
\usepackage{subfigure}
\usepackage[sort&compress]{natbib}
\usepackage{graphicx}
\usepackage{psfrag}
\usepackage{amsmath}
\usepackage[usenames]{color}

\newcommand{\gpfiglarge}[1]{%
  \resizebox{0.45\textwidth}{!}{\includegraphics{#1}}\\[\baselineskip]
  }

\newcommand\rv{{\bf r}}

\newcommand\tauperp{\tau_{\perp}}

\newcommand\tauroddil{\tau_{rod}^{0}}
\newcommand\taue{\tau_{e}}
\newcommand\taurep{\tau_{rep}}
\newcommand\zetaperp{\zeta_{\perp}}
\newcommand\zetapar{\zeta_{\parallel}}

\newcommand\Lp{L_{p}}

\newcommand\msdmidbeadperp{\langle \Delta \rv_{m,\perp}^{2} (t) \rangle}

\newcommand\msdtubedia{\langle \Delta d^{2} (t) \rangle}

\newcommand\Gtot{G(t)}
\newcommand\Gcurv{G_{\mathsf{curv}}(t)}
\newcommand\Gornt{G_{\mathsf{ornt}}(t)}
\newcommand\Gtens{G_{\mathsf{tens}}(t)}
\newcommand\Gcurvplateau{G_{\mathsf{curv,0}}}
\newcommand\Gtotplateau{G_{\mathsf{0}}}

\newcommand\taurod{\tau_{rod}}

\begin{document}

\title{Chain motion and viscoelasticity in highly entangled solutions
  of semiflexible rods}

\author{Shriram Ramanathan}
\affiliation{Department of Chemical Engineering and Materials Science,
  University of Minnesota, Minneapolis, MN 55455, USA}
\author{David C Morse}
\email{morse@cems.umn.edu}
\affiliation{Department of Chemical Engineering and Materials Science,
  University of Minnesota, Minneapolis, MN 55455, USA}

\date{\today}

\begin{abstract}
Brownian dynamics simulations are used to study highly entangled 
solutions of semiflexible polymers. Bending fluctuations of 
semiflexible rods are signficantly affected by entanglement only 
above a concentration $c^{**}$, where $c^{**}\sim 10^{3}L^{-3}$ 
for chains of similar length $L$ and persistence length. For 
$c > c^{**}$, the tube radius $R_{e}$ approaches a dependence
$R_{e} \propto c^{-3/5}$, and the linear viscoelastic response 
develops an elastic contribution that is absent for $c < c^{**}$. 
Experiments on isotropic solutions of $F$-actin span 
concentrations near $c^{**}$ for which the predicted asymptotic 
scaling of the plateau modulus $G \propto c^{7/5}$ is not yet 
valid. 
\end{abstract}

\date{\today}
\maketitle

Solutions of long polymers become entangled when the concentration 
or chain length exceeds a threshhold.  The nature of 
``entanglement" is obviously different, however, for random walks,
rigid rods, and semiflexible threads. It has been proposed that 
solutions of semiflexible rods, of length $L$ less than or equal
to their persistence length $L_{p}$, may exhibit two different 
levels of entanglement, in different concentration regimes
\citep{odijk1983,doi1985,semenov1986,morse1998a} 
-- a loosely-entangled regime, in which only rotations and 
transverse translations are hindered by collisions, and a 
tightly-entangled regime, in which transverse shape fluctuations 
are also strongly affected. 
The crossover between these two regimes is expected to be associated 
with a qualitative change in viscoelastic properties, due to the 
inability of a tightly-entangled solution to rapidly relax stress 
arising from transverse chain deformations.
Clear experimental evidence of ``tight'' entanglement has been 
obtained only for solutions of very long actin protein filaments 
(F-actin), of length $L \sim L_{p} \sim 10 \mu{\rm m}$ and diameter 
$d \sim 10$ nm. The evidence comes both from visualization of 
flourescently labelled chains \citep{kasetal1994,kasetal1996} and 
from rheological measurements 
\citep{satoetal1985,hinneretal1998,schmidtetal2000a,gardeletal2003}.  
It remains unclear, however, whether bending fluctuations are ever
significantly hindered in isotropic solutions of any of a variety 
of other well-studied model systems of semiflexible rods 
with $L \sim \Lp$ \cite{SatoTeramoto1996},
for which the average chain lengths and aspect ratios are all much 
smaller than those obtainable with $F$-actin. Simulations offer 
a potentially important complement to the experimental study of 
these systems, which provide access to different information and 
are subject to different difficulties than those encountered in 
experiments. 

Consider a solution of thin semiflexible rods, each of contour 
length $L$ and persistence length $L_{p}$, with $L \alt L_{p}$.  
Let $c$ be the number density of polymers and $\rho \equiv c L$ 
be the contour length per volume. Simple geometrical arguments
suggest the following sequence of concentration regimes
\cite{morse1998a}:
At dilute concentrations $c < c^{*}$, where $c^{*} \propto L^{-3}$, 
chain motion is essentially unhindered. In the loosely-entangled 
regime, $c^{*} \ll c \ll c^{**}$, rotations and transverse rigid 
body translations are strongly hindered, but tranvserse bending 
fluctuations are not.  In this regime, each chain is trapped in 
a cylindrical cage or tube of radius $R_{e} \sim 1/(cL^{2})$ 
\cite{doi1975}. Above a threshhold 
$c^{**} \sim \sqrt{L_{p}/L}\;c^{*}$, 
this cage become narrow enough to also hinder thermal
bending fluctuations \citep{doi1985,odijk1983,semenov1986,morse1998a}. 
At concentrations $c \gg c^{**}$, chain motion can be described 
by a modified tube model \citep{doi1985,semenov1986,morse1998b} in 
which each chain undergoes reptation in a narrow wormlike tube. 
A scaling argument due to Odijk and Semenov 
\cite{odijk1983,semenov1986,morse1998a} predicts a tube radius
$R_{e} \propto L_{p}(\rho L_{p}^{2})^{-3/5}$ for $c \gg c^{**}$

Our simulations use a novel algorithm that was designed to allow 
simulation of Brownian dynamics of arbitrarily thin but uncrossable 
wormlike threads.
Each polymer is represented as a discretized chain of $N$ inextensible 
rods and $N+1$ beads. A periodic cubic simulation cell is initially 
populated with a thermally equilibrated solution of wormlike chains, 
by a Monte Carlo growth algorithm. At each step of our dynamical 
simulation, a trial move is generated for a randomly chosen chain by 
taking one time step of the Brownian dynamics (BD) algorithm used in 
previous work on dilute solutions 
\citep{pasquali2001,Shankar2002,pasquali2005}. 
A trial move is rejected, however, if it would cause the chosen chain 
to cut through any other. Whether or not a move is accepted, another 
chain is then chosen at random, and the process is repeated. In this
work, we use an algorithm for chains with anisotropic friction, with 
$\zetapar/\zetaperp=1/2$ \cite{pasquali2005}, where $\zetapar$ and 
$\zetaperp$ are longitudinal and transverse friction coefficients, 
respectively. Details of the algorithm are presented elsewhere 
\citep{shriram-thesis,ramanathan2006b}.  
Here, we present results for entangled solutions of chains with 
$L/L_{p} = 0.25 - 2.0$ and $N=10-40$ rods at concentrations 
$ cL^{3} = 0 - 4000$. A 1 mg/ml solution of (hypothetically) 
monodisperse F-actin filaments with $L=8 \mu$m would have $cL^{3} 
\simeq 2500$.

To characterize the effect of entanglement upon bending fluctuations, 
we have calculated two measures of the transverse mean-squared 
displacement (MSD) vs. time for the middle bead of a polymer. 
The quantity $\msdtubedia$, shown in the main plot in Figure 
\ref{fig:msdtubediaLLp}, is the variance of the distance 
$\Delta d(t)$ between the middle bead at time $t$ and the closest 
point on the contour of the same chain at an earlier time $t=0$. 
The inset shows $\msdmidbeadperp \equiv 
\langle |\rv_{m,\perp}(t) - \rv_{m,\perp}(0)|^{2} \rangle$, in 
which $\rv_{m,\perp}(t)$ is the transverse component (transverse 
to the local chain tangent) of the displacement of the middle bead 
from the chain's center of mass. The quantity $\msdmidbeadperp$ is 
not sensitive to center-of-mass diffusion, but only to displacements 
arising from bending fluctuations, and so approaches a finite value 
at long times. 
\begin{figure}
  \begin{center}
    \gpfiglarge{figs/Fig1.eps}
    \vspace{-0.5cm}
    \caption{$\msdtubedia / L^{2}$ vs $t / \tauroddil$ for chains with 
    $L = L_{p}$, where $\tauroddil \equiv \zeta_{\perp}L^{3}/(72k_{B}T)$ 
    is the rod rotation time in dilute soluton.  Numbers near curves are 
    values of $cL^{3}$ (or 'Dilute' for $c=0$), while numbers in 
    parentheses indicate $N$.  The black dashed line is the predicted 
    asymptote at early times, for which $\msdtubedia \propto t^{3/4}$.
    Inset: $\msdmidbeadperp$ vs $t / \tauroddil$ for $L=L_{p}$ and 
    $cL^{3}=$ 'Dilute', 100, 250, 500, 1000, and 2000.  The red dashed 
    line is the result of a slithering-snake simulation of pure 
    reptation. \cite{ramanathan2006b} }
    \vspace{-0.5cm}
    \label{fig:msdtubediaLLp}
  \end{center}
\end{figure}
At early times, both measures of transverse MSD increase as $t^{3/4}$, 
as predicted \citep{granek1997}. With increasing concentration, both 
quantities become suppressed over a range of intermediate times, 
indicating the formation of a tube.

If each chain were confined to a tube of well-defined radius $R_{e}$ 
over a wide range of intermediate times, $\msdtubedia$ would develop 
a plateau, with a plateau value $\msdtubedia \simeq 4R_{e}^{2}$. Here, 
$R_{e}^{2}$ is defined, as in Ref. \cite{morse2001}, as the variance 
of the transverse displacement of the chain from the ``center'' of 
the tube (i.e., the average chain contour) in either of two transverse 
directions. A plateau could appear in $\msdtubedia$ even in the rigid
rod limit, however, due to suppression of transvserse center-of-mass
motion.  The suppression of $\msdmidbeadperp$ at intermediate times,
however, is evidence of hindered {\it bending} motion, and thus of 
tight entanglement.  In fact, we never observe a clean plateau in 
either quantity. Instead, we see a crossover from $t^{3/4}$ growth 
at small $t$ to a much slower growth at intermediate times, which
becomes flatter with increasing concentration and/or chain length
(i.e., increasing $cL^{3}$), with a crossover time $\taue$ that 
decreases with increasing $c$. The suppression in $\msdmidbeadperp$ 
is signficant only for $cL^{3} \gtrsim 500$, suggesting a crossover 
$c^{**} \simeq 500/L^{3}$ for $L=L_{p}$.  

For $cL^{3}=1000$, our results for $\msdmidbeadperp$ include both
a plateau at intermediate times and an upturn at the end of this
plateau. This upturn is mimicked very accurately by the results 
of a separate slithering-snake simulation of pure reptation of a 
wormlike chain (the red dashed line in the inset) 
\citep{ramanathan2006b}. Pure reptation yields a nonzero transverse 
MSD $\msdmidbeadperp$ at times less than the reptation time 
$\tau_{rep} = \zetapar L^{3}/(\pi^{2} k_{B}T)$ because reptation 
occurs along a curved tube. $\msdtubedia$ is defined so as not to 
be affected by pure reptation, and shows a slightly broader plateau 
than $\msdmidbeadperp$.

\begin{figure}
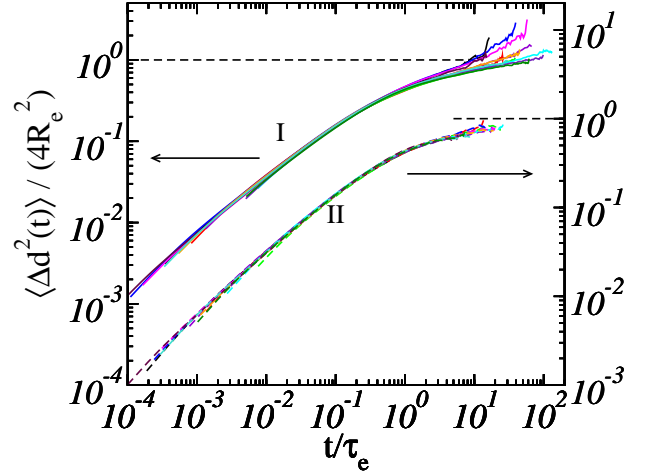

  \begin{center}
    \gpfiglarge{figs/Fig2.eps}
    \vspace{-0.5cm}
    \caption{Collapse of $\msdtubedia$ data.  Curves labelled ``I''
      represent collapse of data for $N=20$, $cL^{3}=250,500,1000$, 
      and $L/L_{p}=0.25,0.5,1.0,2.0$. Curves labelled ``II'' represent 
      collapse of data for $N=40$, $cL^{3}=1000,2000,4000$,
      and $L/L_{p}=0.25,0.5,1.0,2.0$.}
    \vspace{-0.5cm}    
    \label{fig:msdtubediacollapse20-40}
  \end{center}
\end{figure}

To quantify $\tau_{e}$ and $R_{e}$, we have collapsed our data for
$\msdtubedia$ in a manner that assumes the existence of a scaling 
relationship $\msdtubedia=4R_{e}^{2}f(t/\taue)$.  That is, we have 
chosen values for $R_{e}$ and an entanglement time $\tau_{e}$ for 
each set of parameters so as to collapse the data for many different 
values of $L/\Lp$ and $cL^{3}$ onto a master curve of 
$\msdtubedia/(2R_{e})^{2}$ {\it vs.} $t/\tau_{e}$.  The resulting 
collapse is shown in Figure \ref{fig:msdtubediacollapse20-40}. We 
display separate master curves for chains with $N=20$ and $N=40$ 
because early time behavior is noticeably different for discrete
chains with different numbers of rods.
The collapse is excellent for solutions with $cL^{3} \geq 1000$. The 
horizontal dashed lines with $\msdtubedia/4R_{e}^{2}=1$ represent an 
assumed long time asymptote for hypothetical systems of much longer 
chains, from which we have extracted estimates of $R_{e}$. 

\begin{figure}
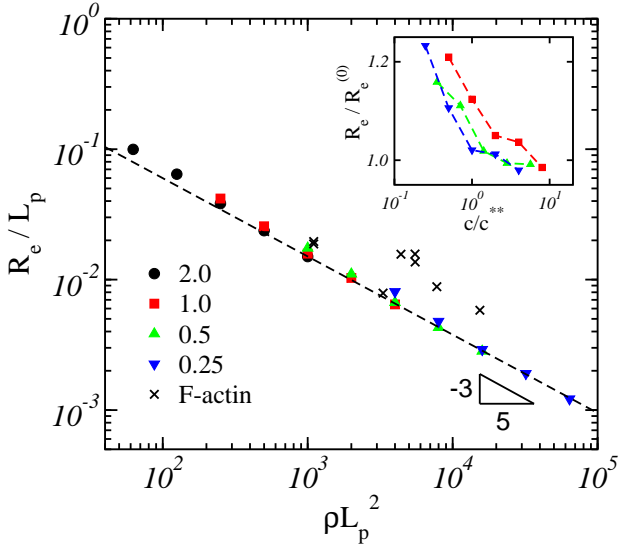

  \begin{center}
    \gpfiglarge{figs/Fig3.eps}
    \vspace{-0.5cm}
    \caption{Non-dimensionalized tube radius vs. concentration.
      Numbers in the legend are values of $L/L_{p}$.  Crosses 
      are flourescence microscopy results for $F$-actin 
      \cite{kasetal1994,kasetal1996}, non-dimensionalized by 
      $\Lp = 17 \mu m$. Inset: Deviation
      $R_{e}/R_{e}^{(0)}$ from the asymptote vs. $c/c^{**}$, 
      where $R_{e}^{(0)} \equiv 0.95 \Lp (\rho \Lp^{2})^{-3/5}$ 
      corresponds to the dashed line in the main plot, and where
      $c^{**} = 500 L_{p}^{1/2} L^{-7/2}$. }
  \vspace{-0.5cm}      
  \label{fig:Re-rho-Lp2-sim-expt}
\end{center}
\end{figure}

Figure \ref{fig:Re-rho-Lp2-sim-expt} shows resulting values of the 
dimensionless tube radius $R_{e}/L_{p}$ vs.dimensionless concentration 
$\rho L_{p}^{2}$ for systems with $L/L_{p}=0.25-2.0$.  
Dimensional analysis requires that the ratio $R_{e}/L_{p}$ be a 
function $R_{e}/L_{p}=f(\rho L_{p}^{2},L/L_{p})$ of dimensionless 
length $L/L_{p}$ and dimensionless concentration $\rho \Lp^{2}$ 
alone. In the tightly-entangled regime, however, we expect $R_{e}$ 
to become independent of $L$, implying that $R_{e}/L_{p}$ must 
approach a function of $\rho L_{p}^{2}$ alone for $c \gg c^{**}$. 
At high concentrations, our results for different values of $L/\Lp$ 
do indeed approach a common asymptote, which is furthermore very 
accurately described by the predicted relation 
$R_{e}/\Lp = \alpha (\rho \Lp^{2})^{-3/5}$, with 
$\alpha = 0.95$ (dashed black line). For each value of $L/\Lp$,
$R_{e}/L_{p}$ also exhibits small but systematic deviations from 
this asymptote at lower concentrations.  This deviation is seen
most clearly in the inset, in which we plot the ratio 
$R_{e}/[0.95 \Lp (\rho\Lp^{2})^{-3/5}]$ vs. $c/c^{**}$, 
where we have taken $c^{**} = 500 \Lp^{1/2}L^{-7/2}$. The near 
collapse of the deviations from the asymptote for the stiffest 
2 chains ($L/\Lp=$0.25 and 0.5) is consistent with the prediction 
that $c^{**} \propto \Lp^{1/2} L^{-7/2}$ for $L \ll \Lp$. 
\citep{morse1998a}. The experimental values for $R_{e}$ in F-actin 
solutions (crosses) are the flouresence microscopy results of 
K\"{a}s {\it et al.} \cite{kasetal1994,kasetal1996}, as defined 
and presented previously in Ref. \cite{morse2001}.

The crossover from loose- to tight-entanglement is expected to 
cause a dramatic change in viscoelastic behavior. Detailed theories 
of linear viscoelasticity have been developed for the extreme limits
of dilute solutions ($c \ll c^{*}$) \cite{pasquali2001,Shankar2002} 
and of very tightly entangled solutions ($c \gg c^{**}$)
\cite{morse1998b}. Both theories make use of a formal  decomposition 
of the stress into curvature, orientational, and tension contributions 
\cite{morse1998a}, and a corresponding decomposition of the dynamic 
modulus $G(t)$ (i.e., the response to an infinitesimal step strain) 
as a sum $\Gtot = \Gcurv  + \Gornt + \Gtens $.  In both dilute and 
loosely-entangled solution, $\Gcurv$ and $\Gtens$ are predicted to 
exhibit power law decays at very early times, but to decay exponentially
at times greater than the relaxation time
$\tauperp = \beta \zetaperp L^{4}/(k_{B}T \Lp)$ of the longest wavelength 
bending mode, where $\beta = (4.74)^{-4}$.  For $c < c^{**}$, $\Gtot$ 
is thus dominated at $t > \tauperp$ by a more slowly decaying orientational 
modulus $\Gornt \simeq (3/5)ck_{B}T e^{-t/\taurod}$, where $\taurod$ is 
a rotational diffusion time.  In loosely-entangled solutions, the only 
predicted effect of entanglement is to increase $\taurod$, without 
significantly changing $\Gcurv$ or $\Gtens$. The plateau of magnitude
$(3/5)ck_{B}T$ in $\Gornt$, which is present even in dilute solution, 
reflects the free energy cost of partially aligning an initially random 
distribution of rod orientations. The crossover to tight entanglement, 
however, is expected to cause a plateau to appear in $\Gcurv$, with a 
plateau value $\Gcurvplateau$ that varies as 
\cite{IsambertMaggs1996,morse1998a,morse1998b}
$\Gcurvplateau \propto k_{B} T \rho^{7/5} L_{p}^{-1/5}$
for $c \gg c^{**}$.

We have ``measured'' $G(t)$ and its components by simulating stress 
relaxation after a rapid, small amplitude uniaxial step extension of 
an initially cubic periodic unit cell. Stress is evaluated using 
the virial tensor, as in previous simulations of dilute solutions 
\cite{pasquali2001,Shankar2002}. Measurements of $G(t)$ in dilute 
solution by this method agree to within statistical errors with 
those obtained previously \cite{pasquali2001,Shankar2002} from stress 
fluctuations in equilibrium. 

\begin{figure}
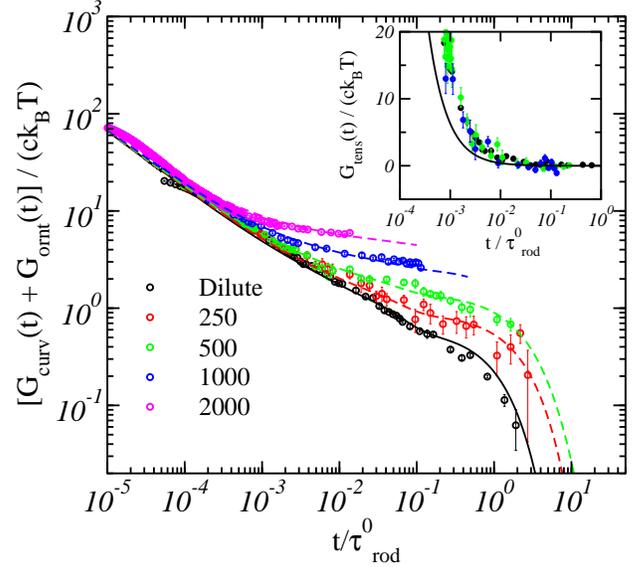

  \begin{center}
  \gpfiglarge{figs/Fig4.eps}
  \vspace{-0.5cm}
  \caption{Non-dimensionalized sum $[\Gornt+\Gcurv]/(ck_{B}T)$ of 
  the orientational and curvature moduli vs $t/\tauroddil$ for 
  $L/L_{p}=0.5$. Numbers in the legend are values of $cL^{3}$. 
  Dashed curves are fits, as discussed in the text.  Black solid 
  curves in both plots are theoretical predictions for dilute 
  solutions \citep{Shankar2002}.  Inset: Corresponding tension 
  stress $\Gtens/(ck_{B}T)$ for systems with $c L^{3}$ = 0, 250, 
  500, 1000.} 
  \vspace{-0.5cm}
  \label{fig:G-curv-all-curve-fits}
  \end{center}
\end{figure}

In Figure \ref{fig:G-curv-all-curve-fits}, the main plot shows a 
non-dimensionalized sum $[\Gornt + \Gcurv]/(ck_{B}T)$ of the two
components of $G(t)$ that are predicted and observed to exhibit 
an elastic plateau. The inset shows $\Gtens/(ck_{B}T)$, which,
as expected \cite{morse1998b}, does not exhibit a plateau, 
and which is found to be almost independent of $c$ over this range 
of parameters.  The plateau in $[\Gornt + \Gcurv]/(ck_{B}T)$ 
becomes significantly greater than the limiting value of 
$3/5$ obtained in dilute solution, which arises from $\Gornt$
alone, only above an apparent crossover concentration of 
$c^{**}L^{3} \sim 250-500$, above which $\Gcurv$ also begins to
contribute to the observed plateau. The terminal relaxation is 
accessed in our simulations only for $cL^{3} \leq 500$, but the 
plateau value is always accessible.

To quantify the plateau modulus, we have fit the sum
$\Gcurv + \Gornt$ to a function
\begin{equation}
  \label{eq:G-curv-sim-curve-fits}
  \frac{3}{5}c k_{B}T e^{-t/\taurod} +
  G_{\mathsf{curv,dil}}(t) +
  \Gcurvplateau e^{-t /\tau_{0}}
  \quad,
\end{equation}
Here, the first term on the r.h.s. is an expression for $\Gornt$, 
where $\taurod$ is a concentration-dependent rotational diffusion 
time, and $G_{\mathsf{curv,dil}}(t)$ is the prediction of 
\citeauthor{Shankar2002} for $\Gcurv$ in dilute solution. The
quantity $\Gcurvplateau$ is the contribution of $\Gcurv$ to the 
overall plateau modulus, which is an adjustable parameter. We 
have used a time constant $\tau_{0} = \taurep/2$ for the 
relaxation of the curvature plateau. This was chosen to fit the 
observed decay of $\Gcurv$ alone (not shown separately here) at 
$cL^{3}=250$ and $500$, and is consistent with a double-reptation 
model of the relaxation of the curvature plateau. Values of the
rotational diffusion time $\taurod(c)$ were measured in separate 
equilibrium simulations \cite{shriram-thesis}, which yield 
$\taurod/\tauroddil = (2.10,3.06,4.46,6.17)$ for 
$cL^{3}=(250,500,1000,2000)$. The values of $\Gcurvplateau$ 
obtained by fitting this data depend very little upon our 
choices for the time constants $\tau_{0}$ and $\taurod(c)$.

The total plateau modulus $G_{0}$ in $\Gtot$ is a sum
$G_{0} = (3/5)ck_{B}T + G_{\mathsf{curv,0}}$ of orientational and
curvature contributions. Figure \ref{fig:G-tot-plateau-sim-actin} 
compares simulation results for $G_{0}$ and $G_{\mathsf{curv,0}}$ 
to reported values of $G_{0}$ in entangled F-actin solutions 
\cite{hinneretal1998,gardeletal2003}. 
\begin{figure}
  \begin{center}
    \gpfiglarge{figs/Fig5.eps}
    \vspace{-0.5cm}
    \caption{$ (\Gtotplateau L_{p}) / (\rho k_{B} T) $ vs $ \rho
    L_{p}^{2} $ from simulations and experiments on entangled
    F-actin solutions. Simulation results for systems with 
    $L/\Lp=0.5$ are shown for both $\Gtotplateau$ (filled squares) 
    and $\Gcurvplateau$ (filled circles). Experimental results for
    $F$-actin solutions of \citet{gardeletal2003} (open circles) 
    and of \citet{hinneretal1998} for filaments of unregulated 
    length (open triangles) and of average length $16\mu m$ 
    regulated by gelsolin (open squares). Dashed line with a 
    slope of $0.4$ is the prediction of the binary collision 
    approximation of Ref. \cite{morse2001}.}
    \vspace{-0.5cm}
    \label{fig:G-tot-plateau-sim-actin}
    \vspace{-0.5cm}
  \end{center}
\end{figure}
The results of Hinner {\it et al.} \cite{hinneretal1998} were 
obtained by macroscopic rheological measurements, while those of 
Gardel {\it et al.} \citep{gardeletal2003} were obtained from 
two-particle micro-rheology. Our results for $\Gtotplateau$ 
agree well with the values of Hinner {\it et al.}, and are well
within the scatter of results reported in the recent literature. 
A fit of our results for $G_{0}$ to a power of $c$ yields 
$\Gtotplateau/c \propto c^{0.7}$.

It is clear from the simulation data, however, that the range 
of concentrations accessed in our simulations, and most of that 
studied experimentally, lies within about one decade of the 
beginning of a broad crossover to tightly entangled behavior, 
below which $\Gcurv$ does not contribute to $G_{0}$.  As a 
result of this proximity to $c^{**}$, $\Gornt$ dominates $G_{0}$ 
over much of this range, while the contribution $\Gcurvplateau$ 
that is actually predicted to vary as $\Gcurvplateau/c \propto 
c^{0.4}$ in the limit $c \gg c^{**}$ increases much more 
rapidly from nearly zero. The results suggest that the very 
rough agreement between the predicted asymptotic behavior of 
$\Gcurvplateau$ for $c \gg c^{**}$ and measurements of $G_{0}$ 
in $F$-actin may be largely fortuitious.

The isotropic-nematic (IN) transition for rodlike polymers 
occurs at a concentration $c_{IN}L^{3} \simeq 4 L/d$. Values 
of $L/d$ for available model systems with $L \lesssim \Lp$ 
other than $F$-actin, such as Fd virus 
\cite{schmidtetal2000b}
($L \simeq 0.9 \mu m$, $\Lp \sim 2 \mu m$, $d \sim 7$ nm) 
and rod-like poly(benzyl glutatmate) 
\cite{SatoTeramoto1996}
($\Lp \sim 0.15 \mu m$ and $d \sim 2$ nm) 
are all at least 10 times smaller than for $F$-actin, for which 
$L/d \sim 10^{3}$. The IN transition in systems with $L/d < 100$ 
occurs at concentrations $c_{IN}L^{3} \lesssim 400$, at 
which $c < c^{**}$.  Our rough estimate of $c^{**} \sim 500 
\Lp^{1/2}L^{-7/2}$ for $L \lesssim \Lp$ implies that a clear
tightly-entangled isotropic regime for semiflexible rods can 
exist only in systems with $L/d \gtrsim 10^{3}$. This is 
consistent with the fact that a clear rheological signature 
of tight entanglement has been observed only in $F$-actin 
solutions. 

Taken as a whole, our results both provide evidence for 
the correctness of a simple scaling theory for the 
asymptotic dependence of tube radius upon concentration 
in tightly-entangled solutions, and (equally importantly) 
clarify the limits of validity that theory, particularly
as applied to rheology. It appears that bending fluctuations 
of rods with $L \sim \Lp$ are signficantly hindered by 
entanglement only under surprisingly stringent conditions. 

This work has been supported by ACS Petroleum Research Fund 
grant 38020-AC7, using computer resources provided by the 
Minnesota Supercomputer Center and the Univ. of Minnesota 
NSF MRSEC.

\bibliography{polymer} \bibliographystyle{apsrev}

\end{document}